\documentclass[twocolumn,english,aps,superscriptaddress,prb,floatfix,nofootinbib]{revtex4-2}
\usepackage[latin9]{inputenc}
\usepackage{color}
\usepackage{babel}
\usepackage{amsmath}
\usepackage{amssymb}
\usepackage{physics}
\usepackage{mathtools}
\usepackage{graphicx}
\usepackage{hyperref}

\graphicspath{{./}{./figs/}}

\begin{document}
\title{Robust semiclassical magnetization plateau in the kagome lattice}

\author{Gabriel Capelo}
\affiliation{Instituto de F\'isica, Universidade de S\~ao Paulo,  05315-970, S\~ao Paulo, SP, Brazil}

\author{Eric C. Andrade}
\affiliation{Instituto de F\'isica, Universidade de S\~ao Paulo,  05315-970, S\~ao Paulo, SP, Brazil}

\begin{abstract}

Inspired by recent observations of the \(1/3\) magnetization plateau in kagome-based magnets, we investigate the \(J_1-J_2\) Heisenberg model on the kagome lattice under the influence of an external magnetic field.  Although the classical ground state at zero field depends on the sign of \(J_2\),  we find a robust \(1/3\) semiclassical magnetization plateau in both cases. The mechanism that stabilizes this plateau is analogous to that observed in the triangular lattice, where quantum fluctuations select a collinear state from the degenerate classical manifold. We calculate the plateau width, which shows a weak dependence on \(J_2\), using nonlinear spin-wave theory. Additionally, we find that a straightforward approach based on linear spin-wave yields quantitatively accurate results even for \(S=1/2\). Furthermore, we identify a magnetization jump at the saturation field when \(J_2=0\); this jump is related to the presence of a flat band and disappears for  \(J_2 \neq 0\). Our study demonstrates that a semiclassical approach effectively captures the \(1/3\) plateau in the kagome lattice and serves as a valuable tool for interpreting experimental and numerical results alike.
\end{abstract}
\date{\today}
\maketitle


\section{Introduction}

Frustrated quantum magnets have long served as a fertile ground for discovering exotic phases of matter, where the delicate interplay of competing interactions, geometric frustration, and quantum fluctuations can destabilize conventional magnetic order,  giving rise to novel magnetic states ~\cite{lacroix11}.   Over the last three decades,  both theoretical and experimental research have explored the behavior of frustrated magnetic insulators under external magnetic fields.  Due to frustration-enhanced quantum fluctuations,  these systems often exhibit anomalies in their magnetization curves.  A significant feature in this context is the emergence of magnetization plateaus --- regions where the magnetization $m$ remains fixed at a rational fraction of its saturation value over a finite range of magnetic fields \cite{chubukov91,  zhitomirsky00,  honecker04,  alicea09,  ye17},  in manifestation of the order by disorder phenomena ~\cite{villain80,  shender82,  henley89}.


The kagome Heisenberg antiferromagnet is an ideal model system for exploring the novel effects of frustration.  Already in the classical limit,  $S\to \infty$,  where $S$ is the spin size,  its behavior is non-trivial,  as no long-range order is present even at $T \to 0$.  For the opposite limit of $S=1/2$,  its ground state is widely believed to be a quantum spin liquid (QSL).  Studying the different states emerging as a function of $S$ increases our understanding of how quantum behavior emerges or fades in interacting systems,  potentially guiding the discovery of new exotic quantum phases and materials ~\cite{chalker92,chubukov92,harris92,hastings00,singh07,ran07,yan11,messio11,messio12,iqbal13,gong16,he17,lauchli19}.  In the presence of an external magnetic field,  the system with $S=1/2$ shows an interesting evolution of phases.  At low fields,  first $1/9$ magnetization plateau emerges out of the magnetically disordered state  ~\cite{nishimoto13,  picot16,fang23, he24},  followed then by a $1/3$ plateau ~\cite{nishimoto13,capponi13, picot16,okuma19}.  At the saturation field,  a macroscopic magnetization jump occurs,  consistent with a picture of non-interacting magnons ~\cite{schulenburg02,zhitomirsky04}.  Notably,  the $1/3$ plateau is present for all values of $S$ ~\cite{nakano15, picot16,  zhitomirsky02, gvozdikova11} as well as in the presence of perturbations ~\cite{morita23},   and it has been experimentally observed in several kagome-based materials  ~\cite{ishikawa15,goto16,okuma20,kermarrec21,suetsugu24,kato24,jeon24,haraguchi25}.


Motivated by the robustness of the $1/3$ plateau, we present here a comprehensive theoretical study of the spin-$S$ $J_1-J_2$ Heisenberg model on the kagome lattice in the presence of an external field, combining analytical arguments, classical Monte Carlo simulations, and spin-wave theory.  Our motivation is to investigate the effects of perturbations to the nearest-neighbor model,  which are inevitably present in real materials,  and that might drastically alter the nature of both the ground state and the magnetization process.  While the nearest-neighbor model ($J_2 = 0$) is well-known for its quantum disordered ground state,  the introduction of a next-nearest-neighbor coupling $J_2$  stabilizes ordered phases at zero field that evolve continuously into the $1/3$ plateau under an applied magnetic field,  similar to what is observed in the triangular lattice.  We find that the extent of this plateau exhibits a striking robustness against variations in $J_2$~\cite{morita23}.  For $J_2=0$, we also encounter a magnetization jump at the transition to the polarized phase for finite $S$.  This jump,  however,  is unstable for $J_2 \neq 0$.   Our findings reveal how frustration and competing exchange interactions conspire to sustain the $1/3$ plateau,  adding a semiclassical perspective for this phase.

The paper is organized as follows: In Section ~\ref{sec:j1j2},  we introduce the $J_1$-$J_2$ Heisenberg model and analyze its classical ground-state phase diagram, identifying the relevant ordered states and their accidental degeneracies in the presence of an external field.  Section ~\ref{sec:thermal} investigates the lifting of these accidental degeneracies due to thermal fluctuations and the stabilization of the $1/3$ magnetization plateau at finite temperatures.  In Section ~\ref{sec:quantum}  we extend our analysis to quantum fluctuations through a nonlinear spin-wave treatment of the plateau phase,  assessing its robustness in the quantum regime.  Section ~\ref{sec:variational} discusses the magnetization jump at the critical field separating the ordered states from the polarized state for $J_2=0$ .  Finally, in Section ~\ref{sec:conclusion},  we synthesize our results, discussing their implications for ongoing theoretical and experimental studies of frustrated magnetism in kagome systems.  In Appendix~\ref {sec:app_nlsw} we discuss technical details of our spin-wave calculation.

\section{\label{sec:j1j2}Ground state of the $J_1-J_2$ model}

\begin{figure}[t]
\centering{}\includegraphics[width=1\columnwidth]{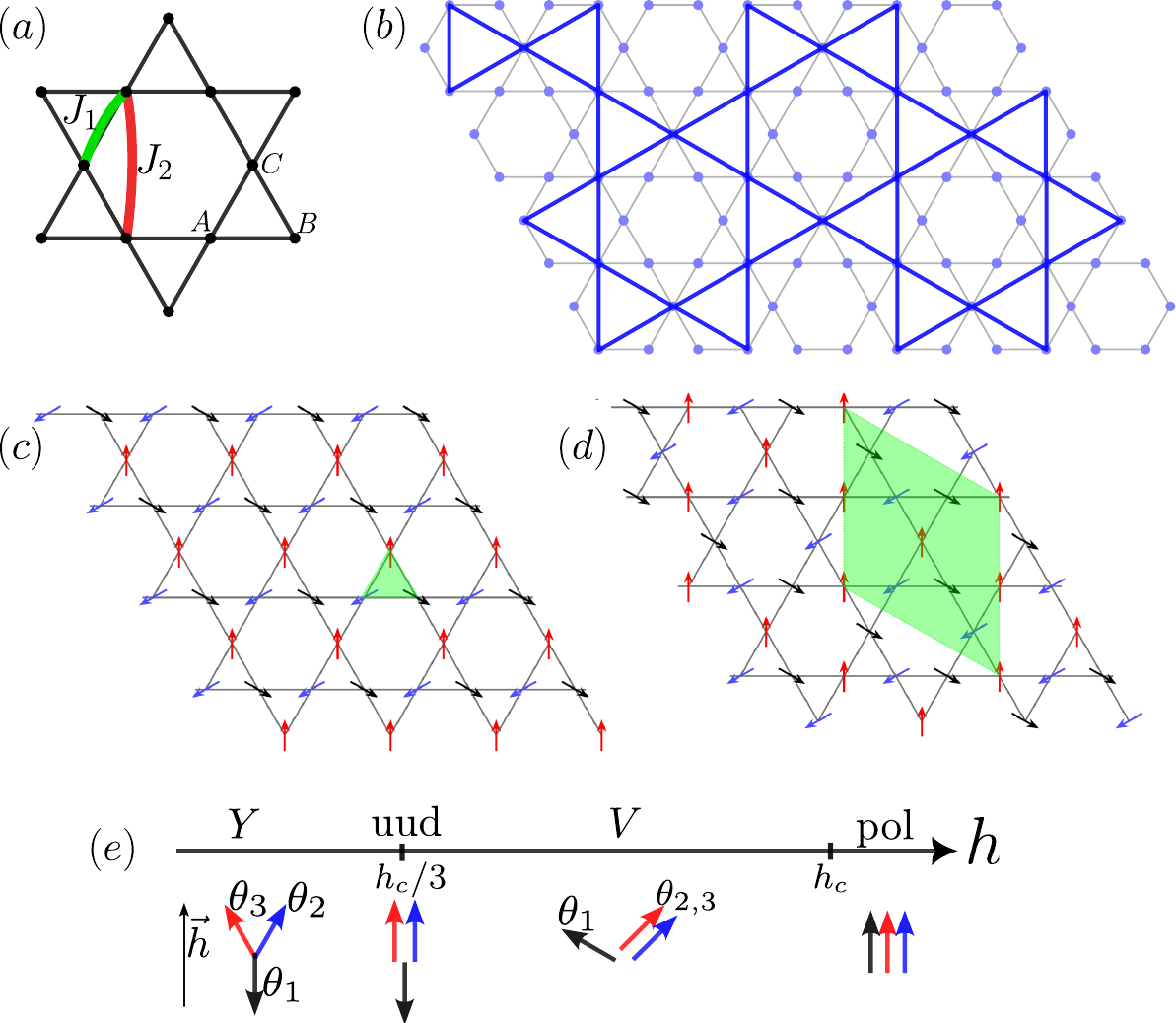}\caption{\label{fig:kag_states0}(a)Kagome lattice showing its three sublattices,  $A$,  $B$,  and $C$. Here, $J_1$ is the nearest-neighbor exchange coupling and $J_2$ the next-nearest neighbor; (b) Network connecting next-nearest neighbours forming a kagome supperlattice;  Semiclassical ground states of the $J_1-J_2$ model: (c) $\vec{Q}=0$ for $J_2>0$,  and (d) $\sqrt{3} \times \sqrt{3}$ for $J_2 < 0$.  The shaded green area indicates the magnetic unit cell; (e) Possible semiclassical states of the $J_1-J_2$ Heisenberg model in a kagome lattice for a finite field.  The three colors represent the three different spins on an elementary triangle. From left to right: Y state,  up-up-down (uud) state, V state,  and polarized state.}
\label{fig:kag_statesh}
\end{figure}

We study the $J_1-J_2$ Heisenberg model in the kagome lattice in the presence of an external field
\begin{equation}
\mathcal{H}=J_1\sum_{\left< ij \right>} \, \vec{S}_i\cdot\vec{S}_j + J_2\sum_{\left<\left< ij \right>\right>} \, \vec{S}_i\cdot\vec{S}_j - h \sum_{i} S^z_i,\label{eq:j1j2}
\end{equation}
where $J_{1\left(2\right)}$ is the coupling between (next-)nearest neighbors,  Fig. ~\ref{fig:kag_states0}(a),  $\vec{S}_i$ is the spin at site $i$,  and $h$ is the external magnetic field.  For convenience, we absorb all the constants that appear in the effective moment $g\mu_{\mathrm{B}}\vec{S}$ of each spin into the field $\vec{h}:=g\mu_{\mathrm{B}}\mu_{0}\vec{H}$.  In this work,  we will focus solely on an antiferromagnetic nearest-neighbor coupling,  and set $J_1 =1$.   

We start the discussion in the classical limit,  $S \to \infty$,   where we treat the spins as a three-dimensional vector of size $S$.  For $J_2=0$, the ground state is highly degenerate, and the system displays no long-range magnetic order ~\cite{chalker92,chubukov92,harris92}.  This stems from the fact that the energy is minimized by any spin configuration for which the total spin of each elementary triangle on the lattice is zero.  However,  this local constraint alone is insufficient to impose long-range order.  To illustrate this fact,  let us restrict ourselves to coplanar spin textures,  which will be the relevant configurations throughout our discussions (see Sec. ~ \ref{sec:thermal} and ~\ref{sec:quantum}).  In this case,  the spins on a triangular plaquette are fixed on a $120^{\circ}$ structure,  with left or right chirality,  see Fig. ~\ref{fig:kag_states0}(c) and (d).  If we associate each spin with a different color, the ground state corresponds to covering the lattice so that every triangular plaquette contains all three colors.  This problem is equivalent to the antiferromagnetic three-state Potts model and thus displays a massive ground-state degeneracy of $S/N=0.1264k_B$ ~\cite{baxter70, huse92}.  This value is a lower bound to the entropy, as noncoplanar ground states are also contained in the classical manifold.

To understand the effect of $J_2$,  at the semiclassical level,  we first notice that this coupling defines three independent kagome superlattices, Fig. ~\ref{fig:kag_states0}(b).   For $J_2>0$,  we need to repeat the $J_1$ constraint on the larger triangles of the superlattice leading to the $\vec{Q}=0$ state,  Fig. ~\ref{fig:kag_states0}(c). Suppose $J_2<0$, a ferromagnetic order is favored inside each of the three superlattices,  while satisfying the constraint imposed by $J_1$ on the elementary triangles of the original kagome lattice,   thus stabilizing the coplanar $\sqrt{3} \times \sqrt{3}$ structure,  Fig. ~\ref{fig:kag_states0}(d).  We find no other phases for larger (absolute) values of $J_2$,  but work in the region of $J_2 < J_1$,  as our goal is to gauge potentially relevant perturbations to the nearest-neighbor Heisenberg model to describe the kagome-based magnets.  For $S=1/2$,  this selection remains true,  but a finite value of $J_2$ is necessary to stabilize the ordered phases, since close to $J_2=0$ we have a quantum disordered state ~\cite{harris92,  iqbal15,  iqbal21}.

In the presence of an external field,  the situation changes because an accidental degeneracy emerges in the problem.  To illustrate, start with the limit $J_2 \to 0$.  It is easy to see that the constraint obeyed by a ground-state configuration is that the total spin of each elementary triangle on the lattice is equal to the $\vec{h}/2J_1$.  This condition gives three equations, while we need to determine six angles for the three classical vectors, leading to accidental degeneracy.  A ferromagnetic $J_2$ has an effect similar to the zero-field case, as it only forces ferromagnetic alignment within each superlattice.  Thus, it does not lift the accidental degeneracy.  If $J_2>0$, we reproduce the accidental degeneracy at the level of the superlattice, again showing that the $J_2$ is incapable of lifting it.  As we show in the following sections,  thermal and quantum fluctuations lift this extensive degeneracy in favor of the two coplanar (Y and V states)  and one collinear  (uud)  spin configurations \cite{zhitomirsky02, schick:tel-03813727,  gvozdikova11},  Fig.  ~\ref{fig:kag_statesh}(e),  similarly to the situation observed in the triangular lattice.  The ground-state energy varies continuously as $E_{\text{cl}}/NS^2=-h^{\text{loc}}/2S-h^{2}/2Sh_c$ with the spin states evolving as shown in Fig.  ~\ref{fig:kag_statesh}. Here, $h^{\text{loc}}/S=2(J_1+J_2)$ for $\vec{Q}=0$, and $h^{\text{loc}}/S=2(J_1-2J_2)$ for the $\sqrt{3}\times\sqrt{3}$ state. The critical field $h_c$ above which the system becomes polarized is $h_c/S=6J_1$ for $J_2 \le 0$ and $h_c/S=6\left(J_1 + J_2\right)$ for $J_2 > 0$.

\begin{figure*}[t]
\centering{}\includegraphics[width=2\columnwidth]{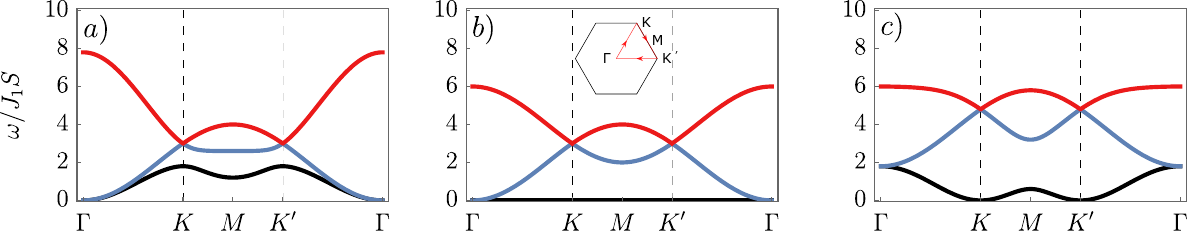}\caption{\label{fig:pol_magnons} Magnon dispersions $\omega_{\textbf{k}}$,  obtained with the linear spin-wave theory,  for the $J_1-J_2$ Heisenberg model in the kagome lattice inside the polarized phase at $h=h_c$.  The dispersions are plotted along the indicated path in the Brillouin zone for different values of $J_2$:  (a) $J_2<0$,  (b) $J_2=0$,  and (c) $J_2>0$.  }
\end{figure*}

We begin by confirming this picture by investigating the instability of the magnon spectrum in the polarized phase.  Performing a linear spin-wave calculation in the saturated regime,  $h \ge h_c$,   (the details of the calculation are presented in Sec.  ~\ref{sec:quantum}),  we obtain the spectrum displayed in Fig.  ~\ref{fig:pol_magnons} at $h=h_c$.  As expected, for $J_2\neq 0$ the magnon gap closes at the transition to the ordered phase.  The wave vector at which this softening occurs indicates the ordering wave vector of the ground-state spin configuration.   For $J_2 = 0$, we obtain the expected flat band for a nearest-neighbor model on the kagome lattice, indicating the absence of an ordered magnetic state.  A finite $J_2$ introduces dispersion to this flat band, and the magnon bands acquire well-defined minima.  They close at the $\Gamma(K)$ point for positive (negative) $J_2$.  These wavectors are compatible with the zero-field ground state shown in Fig.  \ref{fig:kag_states0}.

\section{\label{sec:thermal}Thermal fluctuation and Monte Carlo results}

In the case of the $S = 1/2$ nearest-neighbor kagome lattice Heisenberg antiferromagnet, certain magnetization plateaus correspond to VBCs and thus lack classical counterparts.  On the other hand,  some plateaus exhibit magnetization patterns that can be understood through simple classical analogs --- such as arrangements of down-spins embedded in a field-aligned spin background.  Given their ordered nature,  one might assume that a purely semiclassical theory could fully describe them.  In fact,  Ref.  ~\citep{chubukov91} first predicted a $1/3$ magnetization plateau in the triangular lattice Heisenberg antiferromagnet, showing that the uud state is stabilized due to fluctuations,  Fig.  \ref{fig:kag_statesh}(e). 

In the previous section, we demonstrated that the next-nearest-neighbor exchange $J_2$ lifts the massive degeneracy of the $J_2=0$ classical ground-state manifold at zero field. Still, it cannot lift the accidental degeneracy in a finite field.  Since this degeneracy is not associated with a symmetry, different configurations have distinct excitation spectra.   At finite temperatures, the system explores the phase space near the ground state manifold.  Due to the uneven excitation densities,  it can become trapped near certain states,  effectively lifting the zero-temperature degeneracy,  in a manifestation of the phenomenon called thermal order-by-disorder ~\cite{villain80,  shender82,  henley89},  realizing thus the scenario of Ref.  ~\citep{chubukov91} also for the kagome lattice.

The thermal fluctuations effectively induce a biquadratic exchange,   $-K\sum_{\left< ij \right>} \, \left(\vec{S}_i\cdot\vec{S}_j\right)^2$,  favoring maximally collinear states if $K>0$.  Using a real space perturbation theory ~\cite{zhitomirsky15},  we obtain that $K = k_BTS^2 J_1^2/2(h^{\text{loc}})^2$.  Determining the classical ground state of Eq.  \ref{eq:j1j2} in the presence of a biquadratic term produces the sequence of phases displayed in Fig.~\ref {fig:kag_statesh}(e)  for each of the triangles in the kagome lattice,  independently of the sign of $J_2$.  We then have a qualitative understanding of how the collinear uud state is stabilized for a finite field,  in between the states Y and V,  giving rise to a $1/3$ magnetization plateau,  whose width is  $\Delta h_{\text{plateau}}/S=16KS^2$.  This calculation also suggests a weak dependence of $\Delta h_{\text{plateau}}$ on $J_2$.

\begin{figure}[t]
\centering{}\includegraphics[width=1\columnwidth]{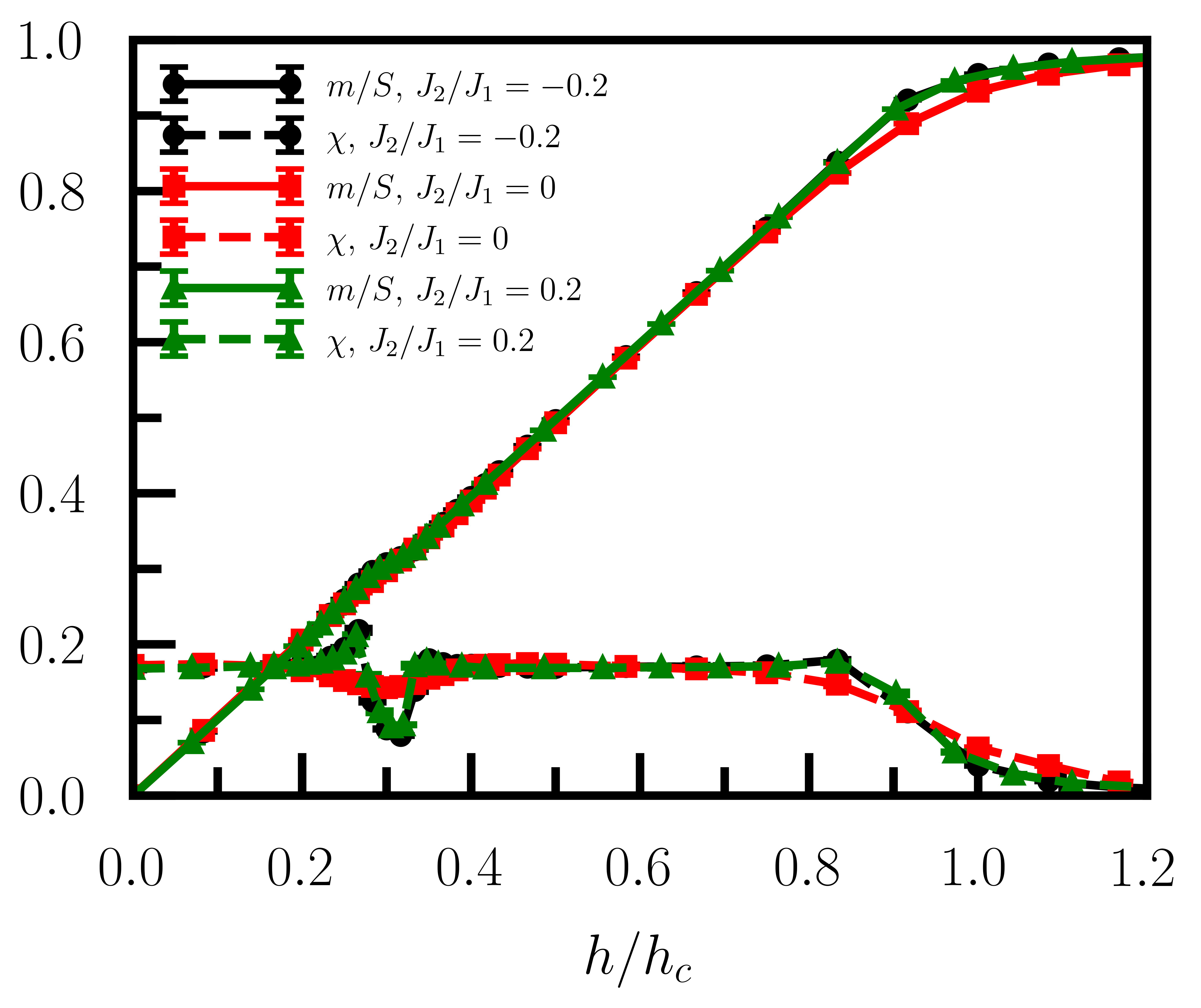}\caption{\label{fig:mag_MC} Classical MC results for the magnetization $m/S$ along the field direction, and its derivative $\chi=dm/dh$,  as a function of the field $h$ for $k_{B}T=0.07J_1$ and $L=24$.   We have $h_c/S=6J_1$ for $J_2\le0$ and $h_c/S=6\left(J_1+J_2\right)$ for $J_2>0$.  For $J_2 > 0$ we renormalized the $\chi$ curve by $\left(J_1 + J_2\right)/J_1 $ to account for the change in $h_c$ due to $J_2$. }
\end{figure}

We confirm that thermal fluctuations stabilize the $1/3$  plateau beyond the perturbative regime.  We study  Eq.~\ref{eq:j1j2} using classical Monte Carlo (MC) simulations on lattices of linear size $L$, with three sites per unit cell,  Fig.~\ref{fig:kag_states0}(a), and periodic boundary conditions. The total number of sites is $N=3L^{2}$, where $L=24$. We perform equilibrium MC simulations using single-site heat-bath and microcanonical updates combined with the parallel tempering method \cite{newman99}.  The results resemble those of the triangular lattice ~\cite{seabra11} and are displayed in Fig. ~\ref{fig:mag_MC}. 

For $T \to 0$,  we have $m=h/h_c$ in the range $h<h_c$,  as expected for an antiferromagnetic state.  For finite $T$,  the magnetization curve evolves smoothly with the field. It shows a $1/3$ plateau near $ h = h_c/3$, which is more evident in the magnetization derivative, which dips in this field range.  The plateau is weak for $J_2=0$,  highlighting its peculiar structure,  as there is no long-range order for any field  ~\cite{zhitomirsky02,  gvozdikova11}.  For finite $J_2$,  at the same temperature,  the plateaus are more pronounced and are surrounded by the usual Y and V phases.  The phase transitions between these phases, as a function of the field, and to the paramagnetic regime as a function of the temperature, are a rich problem ~\cite{zhitomirsky02, seabra11}. Still, we will not pursue it here, as our MC results already confirm that thermal fluctuations lift the accidental degeneracy of the classical ground state and stabilize the $1/3$ plateau.  With the classical phase diagram in hand, in the next section, we will then explore in detail the magnetization process of Eq.~\ref{eq:j1j2} using quantum fluctuations.   

\section{\label{sec:quantum}$1/S$ corrections and quantum fluctuations}

We will discuss in this section the stabilization of the $1/3$ plateau due to quantum fluctuations,  including $1/S$ corrections via a spin-wave calculation.  

\subsection{Linear Spin-wave theory}

Our spin-wave analysis starts with the parametrization of classical phases derived from the Hamiltonian in Eq.~\ref{eq:j1j2}.  In general, each magnetic phase can be described by a unit cell containing $N_s$ spins, requiring the specification of $N_s$ angles $\theta_\mu$,  with $\mu \in \{1,\ldots,N_s\}$ indexing the sites within the magnetic unit cell.  We require a single angle per spin, since the studied classical phases are all coplanar (see Fig. ~\ref{fig:kag_states0}).  The complete set of parametrization angles $\{\boldsymbol{\theta}\} \equiv \{\theta_1,\ldots,\theta_{N_s}\}$ is determined by minimizing the classical ground-state energy of Eq.~\ref{eq:j1j2}.

Given the set of angles, we define local axes $\{\hat{\textbf{x}}_i^{\prime},\hat{\textbf{y}}_i^{\prime},\hat{\textbf{z}}_i^{\prime}\}$ at each site in the following way: $\hat{z}_i^{\prime}$ is parallel to the spin $\textbf{S}_i$ of the classical phase, $\hat{x_i}^{\prime}$ lies the same plane of the spins and $\hat{\textbf{y}}_i^{\prime}=\hat{\textbf{x}}_i^{\prime}\times\hat{\textbf{z}}^{\prime}_i$. Then, we define new spin operators in that base $\textbf{S}_i^{\prime}$. In these new local bases, we apply Holstein-Primakoff (HP) transformations~\cite{holstein1940},
\begin{equation}
\begin{split}
    S_{\mu i}^{z\prime}&=S-a^{\dagger}_{\mu i}a_{\mu i},\\
    S^{-\prime}&=\sqrt{2S}a^{\dagger}_{\mu i}\sqrt{1-\frac{a_{\mu i}^{\dagger}a_{\mu i}}{2S}}, \\ S^{+\prime}&=\sqrt{2S}\sqrt{1-\frac{a_{\mu i}^{\dagger}a_{\mu i}}{2S}}a_{\mu i},
\end{split}
\end{equation}
where $a_{\mu i}^{\dagger}$($a_{\mu i}$) are bosonic creation (annihilation) operators.  Expanding the resulting Hamiltonian in powers of $1/\sqrt{S}$,  we have $\mathcal{H}=\sum_{n=0}^{\infty}S^{2-n/2}\mathcal{H}_n$,  where $\mathcal{H}_n$ denotes terms with $n$ bosonic operators. The linear spin-wave theory (LSWT) consists of truncating the series at harmonic order, i.e., terms with at most two operators.  The Hamiltonian in momentum space reads
\begin{equation}
    \mathcal{H} = S^2E_{\rm{gs},0} + S\sum_{\textbf{k}}\sum_{\mu\nu} \bigg[\mathbb{A}^{\mu\nu}_{\textbf{k}} a^{\dagger}_{\textbf{k}\mu}a_{\textbf{k}\nu} + \frac{1}{2}(\mathbb{B}^{\mu\nu}_{\textbf{k}}a^{\dagger}_{\textbf{k}\mu}a^{\dagger}_{-\textbf{k}\nu} + h.c.) \bigg],
\end{equation}
for some coefficients $\mathbb{A}_{\textbf{k}}^{\mu\nu}$, $\mathbb{B}_{\textbf{k}}^{\mu\nu}$,  in general complex numbers.  Because we take as a starting point a phase of minimum classical energy,  terms linear in the bosonic operators are absent. We diagonalize this quadratic problem via a Bogoliubov transformation, consisting of a linear transformation $\mathbb{T}_{\textbf{k}}$ from the operators $a_{\textbf{k}\mu}$ to new bosonic operators $b_{\textbf{k}\mu}$ and a set of eigenvalues $\omega_{\mu\textbf{k}}$ such that the Hamiltonian is written as
    $\mathcal{H} = S^2 E_{\rm{gs},0} + SE_{\rm{gs},1} + S \sum_{\textbf{k}}\sum_{\mu=1}^{N_s}  \omega_{\textbf{k}\mu} b^{\dagger}_{\textbf{k}\mu}b_{\textbf{k}\mu}$, where $E_{\rm{gs},1} = \frac{1}{2}\sum_{\textbf{k}} \big[ \sum_{\mu=1}^{N_s} \omega_{\textbf{k}\mu} -\rm{Tr}(\mathbb{A}_{\textbf{k}})\big]$. We defined $S^2\mathcal{H}_0\equiv S^2E_{\rm{gs},0}$ which is equal to the classical energy $E_{\rm{cl}}$.  At this order,  the ground-state energy is 
\begin{equation}
    E_{\text{gs}}=S^{2}E_{gs,0}+ \frac{S}{2}\sum_{\textbf{k}} \big[ \sum_{\mu=1}^{N_s} \omega_{\textbf{k}\mu} -\rm{Tr}(\mathbb{A}_{\textbf{k}})\big].
    \label{eq:Enegs}
\end{equation}
Further details of the approach are presented in Appendix \ref{sec:app_nlsw}.

\subsection{\label{sec:text_nlsw}Nonlinear spin-wave theory (NLSWT)}

Inclusions of the next order terms in the $1/S$ power series can be done perturbatively. Retaining the terms $\mathcal{H}_3$ and $\mathcal{H}_4$, we use Wick's theorem to cast them into normal ordering.  The $\mathcal{H}_3$ term is generically finite for noncollinear classical states.  First, for the three boson terms, we have $\mathcal{H}_3=:\mathcal{H}_3:+\mathcal{H}_3^{(1)}$. The double colon denotes normal ordering with respect to the $b_{\textbf{k}\mu}$ bosons, and the superindex means the leftovers that are linear in the number of bosons. The normal ordered term always gives a vanishing contribution in first-order perturbation theory,  and thus is not considered here in order to produce a consistent theory ~\cite{chubukov91, zhitomirsky98, coletta12,coletta16,consoli20}.  We then focus on the last term, $\mathcal{H}^{(1)}_3$.

Because of this extra term,  taking as a starting point a phase that minimizes the classical energy no longer results in the absence of terms linear in the number of bosons.  This implies that we must renormalize the reference state, that is, choose a new set of angles $\{\tilde{\boldsymbol{\theta}} \}=\{\tilde{\theta}_1,\dots,\tilde{\theta}_{N_s}\}$ with $\tilde{\theta}_{\mu}=\theta_{\mu} + \delta \theta_{\mu}/S$.  We impose the condition $S^{3/2}\mathcal{H}_1(\tilde{\boldsymbol{\theta}})+S^{1/2}\mathcal{H}_3^{(1)}(\tilde{\boldsymbol{\theta}})=0$. Expanding in $1/S$ and taking the lowest order, this conditions is
\begin{equation}
    \sum_{\mu}\frac{\partial \mathcal{H}_1}{\partial\theta_\mu}\delta\theta_{\mu} + \mathcal{H}_3^{(1)}=0,
\end{equation}
where both terms of the left-hand side are taken at the classical phase $\theta$.  This condition results in a linear system of $N_s$ equations that determine each $\delta\theta_{\mu}$.

For the terms with four bosons we have $\mathcal{H}_4=:\mathcal{H}_4:+:\mathcal{H}_4^{(2)}:+\mathcal{H}_4^{(0)}$. The last term is a constant shift in energy, and the first term on the right-hand side gives a vanishing contribution in first-order perturbation theory.  We thus focus only on $:\mathcal{H}_4^{(2)}:$, the terms with two bosonic operators:
\begin{equation}
    \mathcal{H}_4^{(2)}=\frac{1}{2}\sum_{\textbf{k}}\alpha_{\textbf{k}}^{\dagger}\tilde{\mathbb{M}}_{\textbf{k}}\alpha_{\textbf{k}}-\frac{1}{4}\sum_{\textbf{k}}\text{Tr} \tilde{\mathbb{M}}_{\textbf{k}}
\end{equation}
where $\alpha_{\textbf{k}}\equiv(a_{\textbf{k}1}, \dots a_{\textbf{k}N_s},a_{-\textbf{k}1}^{\dagger},\dots,a_{-\textbf{k},N_s}^{\dagger})^{T}$ and $\tilde{\mathbb{M}}_{\textbf{k}}$ is a matrix with coefficients that depend on the angles $\theta_{\mu}$ and the avarages
\begin{equation}
\begin{split}
    m_{\mu i \nu j}=\langle a_{\mu i}^{\dagger}a_{\nu j}\rangle \quad&, \quad \Delta_{\mu i \nu j}\equiv \langle a_{\mu i}a_{\nu j}\rangle,\\
    n_{\mu}\equiv\langle a^{\dagger}_{\mu i}a_{\mu i} \rangle \quad&, \quad \delta_{\mu}\equiv\langle a_{\mu i}a_{\mu i}\rangle.
\end{split}
\end{equation}

If we define $\ket{\textbf{k}\mu}\equiv b^{\dagger}_{\textbf{k}\mu}\ket{0}$ (where $\ket{0}$ is the Bogoliubov vacuum), then the correction to the spectrum is obtained by $\delta \omega_{\textbf{k}\mu}=\frac{1}{2}\bra{\textbf{k}\mu}:\alpha_{\textbf{k}}^{\dagger}\tilde{\mathbb{M}}_{\textbf{k}}\alpha_{\textbf{k}}:\ket{\textbf{k}\mu}=\Sigma_{\textbf{k}}^{\mu\mu}$, where $\Sigma_{\textbf{k}}\equiv(\mathbb{T}^{-1}_{\textbf{k}})^{\dagger}\tilde{\mathbb{M}}_{\textbf{k}}\mathbb{T}_{\textbf{k}}^{-1}$.  In that case, the new excitation energies are given by $\tilde{\omega}_{\textbf{k}\mu}=S\omega_{\textbf{k}\mu}+\delta \omega_{\textbf{k}\mu}$.


\subsection{Magnetization curve}

\begin{figure}[t]
    \includegraphics[width=1\linewidth]{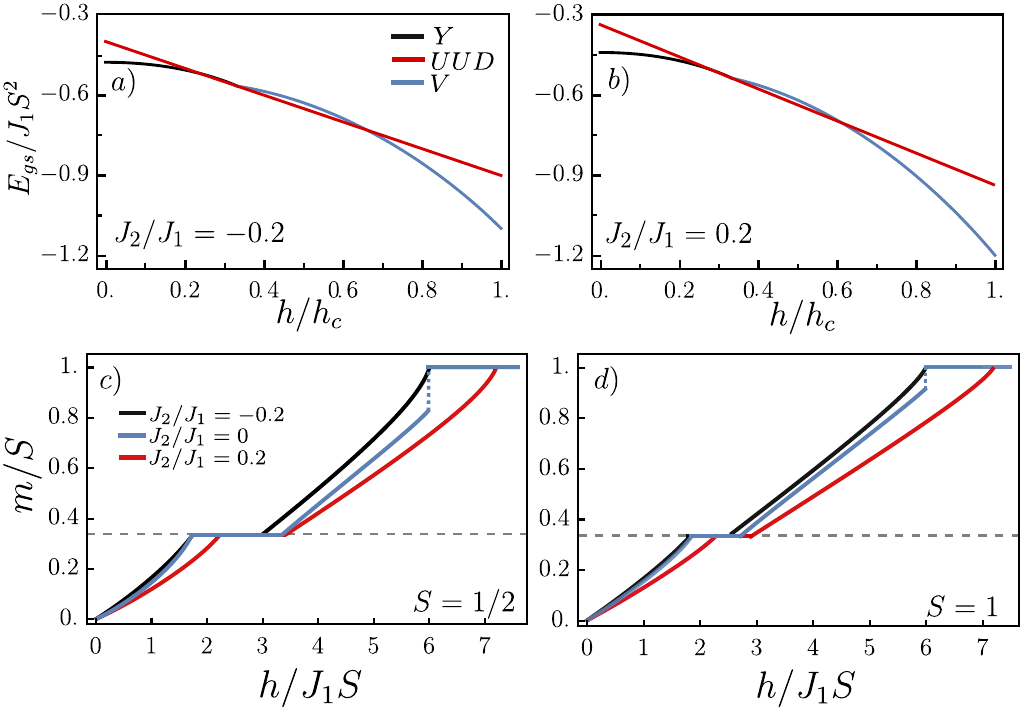}
    \caption{(a) Ground-state energy $E_{\text{gs}}$ as function of the magnetic field $h$ obtained with LSWT, Eq. ~\ref{eq:Enegs}, for $S=1/2$ and $J_2/J_1=-0.2$; The uud energy was calculated with Eq. ~\ref{eq:linextrap}. (b) Same as (a), for $S=1/2$ and $J_2/J_1=0.2$; (c) Magnetization along the field direction $m$ as a function of $h$,  obtained with Eq.~\ref{eq:mh-dEgsdh},  for $S=1/2$ and different values of $J2$: $J_2/J_1=-0.2$, $J_2/J_1=0$, and $J_2/J_1=0.2$; (d) Same as (c) for $S=1$.  The $1/3$ plateau was inserted phenomenologically as discussed in Sec. \ref{sec:MagnetizationPlateau}.}
    \label{fig:EnergiesAndManetizations}
\end{figure}

To establish the effect of quantum fluctuations, we compare the ground-state energy of the uud phase and the neighboring $Y$ and $V$ phases, as shown in Figure ~\ref{fig:EnergiesAndManetizations}.  For the ground-state energies of the $Y$ and $V$ phases, we employ Eq. ~\ref{eq:Enegs}.  The energy of the uud phase can be estimated with this LSWT approach only at $h=h_c/3$,  where it is classically stable,  and we dub it $E^{\mathrm{uud}}_{LSWT}(h_c/3)$.  To extend its validity beyond this field,  we perform the linear extrapolation proposed in Ref. \cite{coletta16},
\begin{equation}
    E^{\mathrm{uud}}(h) = E^{\mathrm{uud}}_{LSWT}(h_c/3) - \bigg(h-\frac{h_c}{3}\bigg)\frac{
    NS}{3}.
    \label{eq:linextrap}
\end{equation}
To justify the second term,  we notice that the total spin projection along the magnetic field is conserved in Eq. ~\ref{eq:j1j2},  leading to ground state energies that depend linearly on the field strength.  The $1/S$ expansion of the Hamiltonian around the uud structure retains this property, even when truncated at harmonic order, as we show shortly.  Because $m/S=1/3$ for the uud state,  we arrive at Eq. ~\ref{eq:linextrap}.  In Figs.~\ref{fig:EnergiesAndManetizations}(a), (b), there is a range of values of the magnetic field where the energy of the uud is the lowest,  in comparison with the energy of the Y and V states,  indicating that the plateau is stabilized in that region.  This proof-of-principle shows that quantum fluctuations lower the energy of the collinear uud phase more than those of the other non-collinear phases, leading to a selection of the plateau phase.  This approach, however, does not produce a continuous magnetization curve with a plateau.

To construct the whole magnetization curve within LSWT for Eq. ~\ref{eq:j1j2},  we first compute the $T=0$ magnetization in the field direction  
\begin{equation}
m =-\frac{1}{N}\frac{\partial E_{\mathrm{gs}}}{\partial h}
=-\frac{S^{2}}{N} \frac{\partial}{\partial h} \left[E_{\mathrm{gs},0} + \frac{E_{\mathrm{gs},1}}{S}+\mathcal{O}\!\left(\frac{1}{S^{2}}\right)\right],
\label{eq:mh-dEgsdh}
\end{equation}
which includes the $1/S$ correction to the ground-state energy in Eq. ~\ref{eq:Enegs}.  To phenomenologically include the plateau,  we cut the resulting curve obtained from Eq. ~\ref{eq:mh-dEgsdh}  at $m/S=1/3$ ~\cite{coletta16},  see Fig. \ref{fig:EnergiesAndManetizations}(c),  (d).  We see that the curves are continuous and show a plateau for all values of $J_2$,  including $J_2=0$.  The plateau width depends weakly on $J_2$ and diminishes with $S$,  as expected.  The magnetization curve for $J_2=0$ also displays a jump at $h_c$,  which we explore in detail in Sec.  \ref{sec:variational}.  

We remark that Eq. ~\ref{eq:mh-dEgsdh} is equivalent to the magnetization obtained from 
\begin{align}
m & =\frac{1}{N}\sum_{i}\left\langle S^{z}_{i}\right\rangle =\frac{S}{N_s}\sum_{\mu}\cos\theta_{\mu}\nonumber \\
 & +\frac{1}{N_s}\sum_{\mu}\left(-\sin\theta_{\mu}\delta\theta_{\mu}+\cos\theta_{\mu} \delta S_{\mu} \right)+\mathcal{O}\!\left(\frac{1}{S}\right),\label{eq:mh <>}
\end{align}
where  $\delta S_{\mu}=\langle a_{\textbf{k}\mu}^{\dagger}a_{\textbf{k}\mu} \rangle$ and the expectation values are calculated with respect to the vacuum of the Bogoliubov quasiparticles \cite{zhitomirsky98,  coletta12}.  The computation here is more involved, as it requires correcting the classical angles due to the cubic terms in the spin-wave Hamiltonian.  In our case,  it is handy to justify the phenomenological approach we employ to construct the magnetization curve~\citep{coletta16}.  

We thus find that a semiclassical approach to solve Eq.  \ref{eq:j1j2} is capable of producing a robust $1/3$ magnetization plateau in the kagome lattice.  In the following,  we present further arguments to confirm the presence of the plateaus and discuss their stability as a function of $S$ and $J_2$.

\subsection{\label{sec:MagnetizationPlateau}Magnetization plateau}

We now explore in further detail the existence of the plateau resorting to NLSWT.  We start with the approach of Ref.  ~\cite{chubukov91} and we compute,  to leading order in $1/S$,  the critical field at which the Y and V phases become collinear,  i.e.,  reduce to the uud state.  We call these fields $h_{1}$ and $h_{2}$,  respectively,  and their difference gives the plateau width.  More precisely, we obtain the $1/S$ correction to the spin angles $\delta \theta_\nu$ in the magnetic unit cell of the Y and V phases.  We then impose the condition $\sum_{\mu}\cos\left(\theta_{\mu}+\delta\theta_{\mu}/S\right)/N_s=1/3$,  which leads to
\begin{equation}
\frac{1}{N_{s}}\sum_{\mu}\left(\cos\theta_{\mu}-\frac{\delta\theta_{\mu}}{S}\sin\theta_{\mu}\right)=\frac{1}{3}.  \label{eq:chubukov_def}
\end{equation}

\begin{figure}[t]
    \centering
    \includegraphics[width=1.05\linewidth]{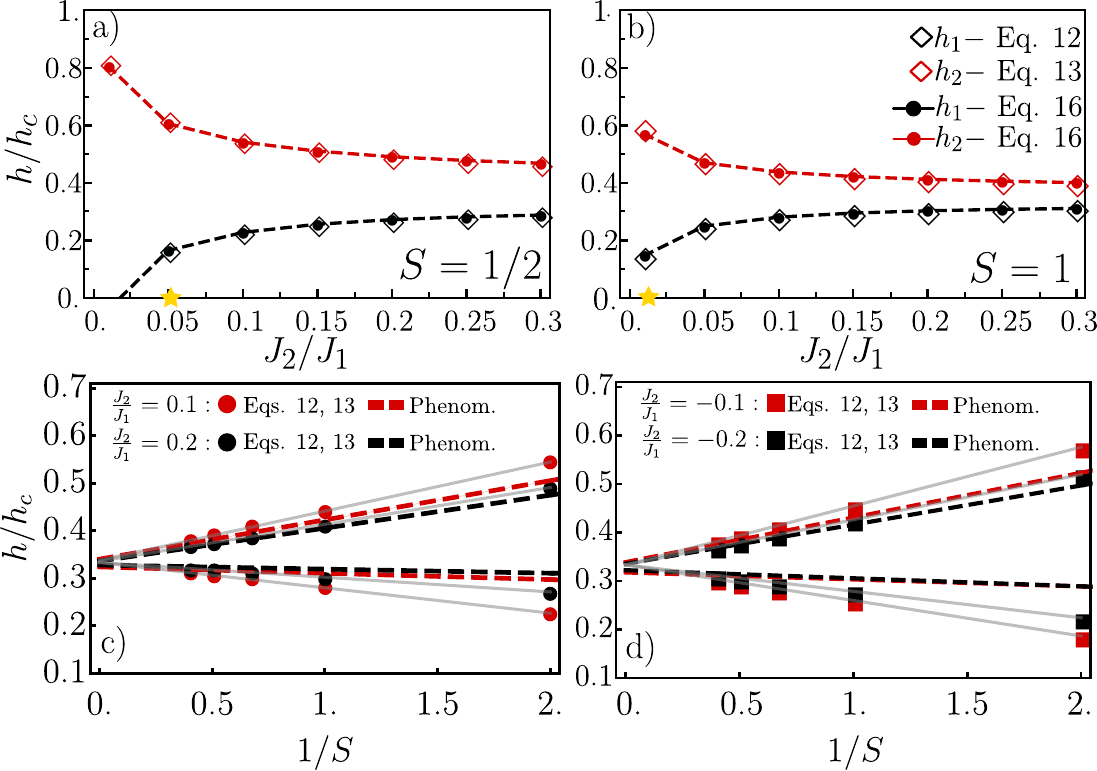}
    \caption{Critical fields $h_1$ and $h_2$ for the $\vec{Q}=0$ state as function of $J_2$ for: a) spin $S=1/2$; and $b)$ spin $S=1$. The open symbols indicate the result of the angle renormalization given by Eqs.~\ref{eq:chubukov1_def}, and ~\ref{eq:chubukov2_def}. The solid symbols with dashed lines represent the result of the gap closing calculated with Eq.~\ref{eq:defhcgap}. The yellow star on the $J_2$ axis indicates the breakdown of LSWT stability for $h=0$, see Sec. \ref{sec:unstable}; Critical fields of the plateau phase as a function of $1/S$ for: c) $J_2\geq 0$; and d) $J_2\leq 0$. The dashed red (black) lines correspond to results with the phenomenological method discussed in Sec. \ref{sec:MagnetizationPlateau}  for $J_2/J_1=\pm0.1 (J_2/J_1=\pm 0.2)$. The solid symbols indicate the result of the angle renormalization given by Eqs.~\ref{eq:chubukov1_def}, and ~\ref{eq:chubukov2_def}. The gray lines following the solid symbols are a guide for the eye.}
    \label{fig:NlSWTcheck}
\end{figure}

In the expression above,  both $\theta_{\mu}$ and $\delta \theta_{\mu}$ are a function of the field $h$.  If we solve it self-consistently,  we include terms to all orders in $1/S$.  To achieve a consistent expansion,  we use the ground state value for $\cos\theta_{\mu}$ $[(1/N_s)\sum_{\mu} \cos(\theta_{\mu})=h/h_c]$ while evaluating both $\sin\theta_{\mu}$ and $\delta \theta_{\mu}$ at $h_c/3$.  As an example, in the $\vec{Q}=0$ state consider the Y phase,  where we have $\theta_1=\pi$ and $\theta_2=\theta=-\theta_3$,  and Eq.  \ref{eq:chubukov_def} reduces to 
\begin{equation}
h_1=\frac{h_c}{3}\left(1+\frac{2}{S}\overline{\sin\theta\delta\theta}\right),
\label{eq:chubukov1_def}
\end{equation}
where the overbar means that the quantities are evaluated in the limit $h\to h_c/3$.  
For the V phase, $\theta_2=\theta_3$, and Eq. \ref{eq:chubukov1_def} reduces to
\begin{equation}
    h_2=\frac{h_c}{3}\left( 1+\frac{1}{S}(\overline{\sin\theta_1\delta\theta_1}+2\overline{\sin\theta_2\delta\theta_2}) \right).
    \label{eq:chubukov2_def}
\end{equation}
In Fig.  ~\ref{fig:NlSWTcheck}(a),(b) we show $h_1$ and $h_2$ as functions of $J_2$.  We see that for $J_2 \gtrsim  0.1 J_1$ the plateau width is weakly dependent on $J_2$,  indicating that the key role of $J_2$ is to stabilize the classical phase,  while $J_1$ governs the plateau emergence.  For smaller values of $J_2$,  quantum fluctuations are strongly enhanced, and the plateau width increases.  As we increase $S$,  the plateau range is also suppressed,  as quantum fluctuations become smaller.  

Ref. ~\cite{chubukov91} also showed that, for the $1/3$ magnetization plateau in the triangular lattice,  the uud structure develops a spin gap --- as expected from the order by disorder mechanism \cite{rau18} --- precisely in the range $h_1 < h <h_2$.  The critical fields where this gap closes coincide with the transition away from collinearity as given by Eq.  ~\ref{eq:chubukov1_def},  setting the boundaries of the uud phase,  see Fig. ~\ref{fig:NlSWTcheck}(a),(b).  

More precisely, we can write the Hamiltonian as
\begin{equation}
    \mathcal{H}=\overline{\mathcal{H}}-\delta h \sum_iS_i^{z}
\end{equation}
where $\delta h=h-h_c/3$. Here we denoted $\overline{\mathcal{H}}$ as the Hamiltonian in Eq. ~\ref{eq:j1j2} taken at $h=h_c/3$. Then, apply HP transformations, with local axes according to the uud state.  For simplicity,  let us discuss the $\vec{Q}=0$ case. Treating $\overline{\mathcal{H}}$ at harmonic order yields three magnon bands,  with two of them containing pseudo-goldstone modes.  Including the effects of interactions via $\overline{\mathcal{H}}_{4}$ ($\overline{\mathcal{H}}_3$ is absent because the phase is collinear),  the three bands $\tilde{\overline{\omega}}_{\textbf{k}\mu}$ become gapped. The gap is obtained by $\overline{\Delta}_{\mu}=\min_{\textbf{k}} \tilde{\overline{\omega}}_{\textbf{k}\mu}$.  Writing $S^{z}_i$ in terms of the $b_{\textbf{k}\mu}$ operators results in
\begin{equation}
    -\delta h \sum_i S^{z}_i=-\delta h \frac{NS}{3} + \delta h\sum_{\textbf{k}}\begin{pmatrix}
        b^{\dagger}_{\textbf{k}1} & b^{\dagger}_{\textbf{k}2} & b^{\dagger}_{\textbf{k}3}
    \end{pmatrix}\Lambda\begin{pmatrix}
        b_{\textbf{k}1}\\
        b_{\textbf{k}2}\\
        b_{\textbf{k}3}
    \end{pmatrix},
    \label{eq:ZeemanAddFieldRotated}
\end{equation}
where $\Lambda=\text{diag}(1,-1,1)$.  The effect of the magnetic field on the spectrum is a uniform linear shift down (up) of the second (first and third) band(s).  The dispersion as function of the field is $\tilde{\overline{\omega}}_{\textbf{k}\mu}\pm\delta h$.  The critical field values are then determined by the closing of the gap, given by $\overline{\Delta}_{\mu}\pm\delta h=0$, or
\begin{equation}
    h_1=h_c/3-\min\{\overline{\Delta}_3,\overline{\Delta}_1\} \quad,\quad h_2=h_c/3+\overline{\Delta}_2.
    \label{eq:defhcgap}
\end{equation}
For example, for $J_2/J_1=0.1$ and $S=1/2$ we have that the three gaps occur at $\textbf{k}=0$ and have the values of $\overline{\Delta}_1=1.1~J_1$, $\overline{\Delta}_2=0.6865~J_1$ and $\overline{\Delta}_3=0.3509~J_1$. The critical fields are then $(h_1,h_2)=(0.7491,1.7864)$, in units of $J_1$.  For the $\sqrt{3}\times\sqrt{3}$ case, we have an analogous result: six of the nine bands shift up by $\delta h$, and the other three shift down by $-\delta h$.  

We checked numerically that the values of the critical fields obtained by Eq. ~\ref{eq:defhcgap} coincide with the values of Eqs. ~\ref{eq:chubukov1_def} and ~\ref{eq:chubukov2_def}, see Fig. ~\ref{fig:NlSWTcheck}.  We thus have established the existence of the plateau, both gauging the instability of the neighboring Y and V phases and the instability of the plateau itself.  These complementary studies produce consistent results to order $1/S$ \cite{chubukov91,consoli20}. 


We are now in a position to justify further the phenomenological approach we are using to construct the full magnetization curve, following the arguments presented in Ref. ~\citep{coletta16} for the triangular lattice. We begin by noting that Eq. ~\ref{eq:mh <>} includes a \(1/S\) correction for both the classical angles and the spin size. However, the condition \(m/S = 1/3\) is equivalent to Eq. ~\ref{eq:chubukov_def} if one consistently works within a \(1/S\) expansion. To establish this equivalence, we recognize that the difference between these two conditions is represented by the term \(\frac{1}{N_s}\sum_{\mu}\cos\theta_{\mu}\delta\theta_{\mu}\). At the consistent order in \(1/S\), this term is equal to \(\sum_{\mu}\overline{\cos\theta_{\mu}\delta S_{\mu}}\). Remarkably, we found that this term vanishes identically, as confirmed through numerical checks. Taking as an example $J_2/J_1=-0.1$ we obtain $\overline{\delta S_1}=\overline{\delta S_2} = \overline{\delta S_3} = 0.3689$ and $\overline{\delta S_4}=\dots=\overline{\delta S_9}=0.1844$, resulting in $-\sum_{\mu=1}^{3}\overline{\delta S_{\mu}}+\sum_{\mu=4}^{9}\overline{\delta S_{\mu}}=0$. The identity holds for all values of $J_2$. This result is rooted in the general fact that,  in the plateau,  $m/S=1/3$ to all orders in $1/S$  because $\sum_i S_i^z$ commutes with $\mathcal{H}$ in Eq.  ~\ref{eq:j1j2}.  

In practice, we set \( m(h)/S = 1/3 \) within the phenomenological approach. This choice results in a small difference in the plateau extent compared to the consistent criteria outlined in Eqs. \ref{eq:chubukov1_def} and \ref{eq:chubukov2_def}, as illustrated in Fig. \ref{fig:NlSWTcheck}(c) and (d). The phenomenological approach tends to underestimate the plateau width, although this discrepancy diminishes as \( S \) increases. Despite its simplicity, this method yields excellent results that can be directly compared to experimental and numerical data.  For example, the phenomenological curves are generally consistent with the exact diagonalization results from Ref. \citep{morita23}, which, for \( J_2 = 0.2 \) and \( T = 0.1 \), found the values \( (h_1, h_2) = (1.2, 1.6) \) in units of \( J_1 \). Our estimate, although based on \( T = 0 \), is \( (h_1, h_2) = (1.12, 1.71) \).  However, this simplified approach cannot capture all features of the magnetization curve. For instance, it produces cusps with finite slopes at the plateau boundaries. At the same time, general theoretical considerations suggest that there should either be a magnetization jump or a continuous transition exhibiting a logarithmic singularity with a divergent slope~\cite{Takano_2011}.

\subsection{\label{sec:unstable} LSWT instability and phase diagram}

\begin{figure*}[t]
    \centering
    \includegraphics[width=1\linewidth]{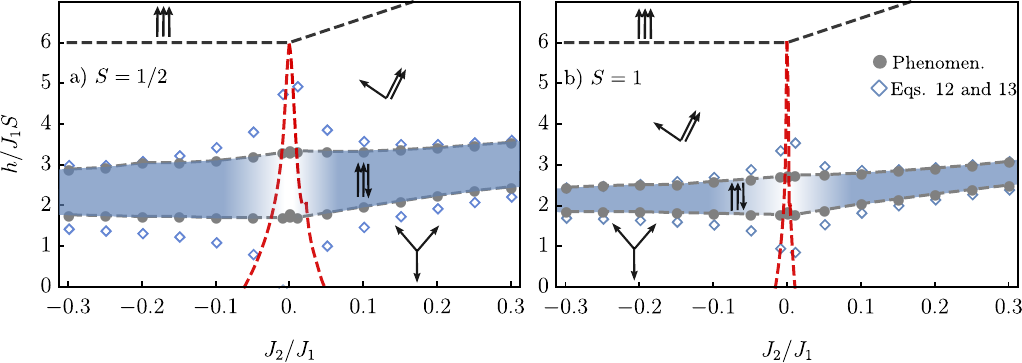}
    \caption{Phase diagram of the model at semiclassical level for (a) $S=1/2$ and (b) $S=1$.  The red dashed lines mark the boundaries of the region of stability of LSWT, defined by points in the parameter space where $\delta S <S$.  Solid circle symbols were obtained using the phenomenological approach, cutting the magnetization curve at $m/S=1/3$.  Open symbols were obtained from the condition imposed on the renormalized angles, yielding critical fields given by Eqs. \ref{eq:chubukov1_def} and \ref{eq:chubukov2_def}. The blue stripe demarcates the uud phase, neighbored by the Y phase (below) and the V phase (above). The polarized phase occurs for $h>h_c$.}
    \label{fig:DiagramaFase}
\end{figure*}

The power series expansion is based on the diluted regime hypothesis, which states that the average $\langle a^{\dagger}_{\mu i} a_{\mu i} \rangle / 2S$ is not large. Although we did not provide a justification for this, we can verify its validity \textit{a posteriori}. We will calculate this explicitly,

\begin{equation}
    \delta S\equiv\frac{1}{N}\sum_{\mu i}\langle a^{\dagger}_{\mu i}a_{\mu i}\rangle=\frac{1}{N}\sum_{\mu \textbf{k}}\langle a_{\mu \textbf{k}}^{\dagger}a_{\mu \textbf{k}}\rangle
\end{equation}
where the average is taken with respect to the Bogoliubov vacuum. As we approach the limit of $J_2 \to 0$, quantum fluctuations become strong, resulting in a divergence of $\delta S$. This means that even for a substantial value of $S$, the LSWT may be poorly justified if we get sufficiently close to $J_2 = 0$, as the corrections to the spin can exceed the spin itself. To ensure stability, we impose the condition $\delta S < S$. For instance, at $h = 0$ and $S = 1/2$, LSWT becomes unstable in the region $-0.06 \lesssim J_2/J_1 \lesssim 0.05$, as shown in Figure ~\ref{fig:DiagramaFase}.  We observe that this instability region diminishes with both $h$ and $S$,  as fluctuations are suppressed. 

The significant impact of quantum fluctuations explains the discrepancy between the plateau widths estimated using the phenomenological method and those obtained from NLSWT.  For values where $|J_2|/J_1 \gtrsim 0.15$, both methods agree reasonably well. However, in the instability region, NLSWT yields an enlarged plateau, leading to a negative $h_1$, which is clearly unphysical.  This discrepancy suggests that the convergence of the $1/S$ power series is not well controlled and indicates a breakdown of magnetic order due to quantum fluctuations as $|J_2| \to 0$.  This region requires a treatment beyond our semiclassical approach.   We have chosen, for visual guidance in Fig. \ref{fig:DiagramaFase}, to use a dashed line that follows the points obtained by the phenomenological method, and we observe that the plateau width depends weakly on $J_2$.  


\section{\label{sec:variational} Magnetization jump}

After discussing the $1/3$ magnetization plateau in the range \( h_1 < h < h_2 \), we turn to another distinct feature of the semiclassical magnetization curve: the jump it exhibits at \( h_c \) when \( J_2 = 0 \) (see Fig. ~\ref{fig:EnergiesAndManetizations}(a), (b)). We construct this curve using Eq. ~\ref{eq:mh-dEgsdh}, considering either the states \( \vec{Q} = 0 \) or \( \sqrt{3} \times \sqrt{3} \) as the reference state. We find that the jump is present for finite spin \( S \) and at zero temperature \( T = 0 \), while it is absent in classical Monte Carlo results (see Fig. ~\ref{fig:mag_MC}).

\begin{figure*}[t]
    \centering
    \includegraphics[width=\linewidth]{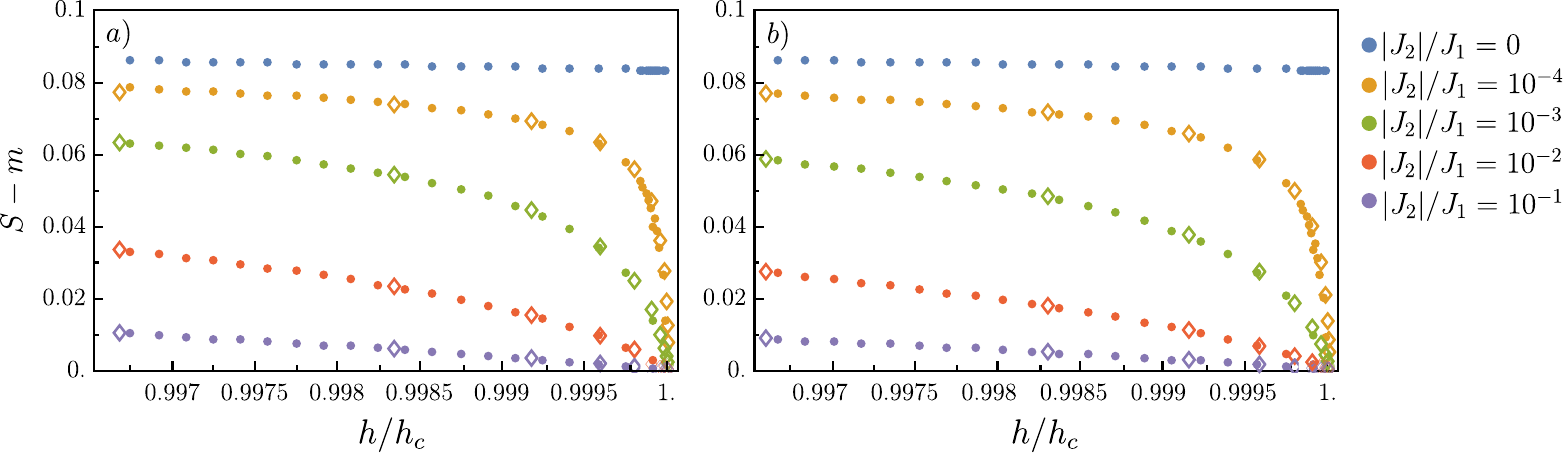}
    \caption{Magnetization $m$ for spin $S=1/2$ close polarization for $a)$ the $\sqrt{3}\times\sqrt{3}\; (J_2\leq 0)$ state; $b)$ $\vec{Q}=0 \;(J_2\geq 0)$ state. The solid circle symbols were obtained using the derivative of the energy with respect to the magnetic field at LSWT order, Eq. ~\ref{eq:mh-dEgsdh}. Open symbols were obtained using Eq. ~\ref{eq:mh <>}.}
    \label{fig:SemiClassJumpCheck}
\end{figure*}

As highlighted in Fig. ~\ref{fig:DiagramaFase}, quantum fluctuations are significant when \( J_2 = 0 \), which is related to the presence of a flat band in this limit (see Fig. ~\ref{fig:pol_magnons}). In Fig. ~\ref{fig:SemiClassJumpCheck}, we show that the jump is indeed absent for finite values of \( J_2 \). We plot \( S - m \) as a function of \( h \) near the polarization field, revealing that this curve approaches zero at \( h_c \). However, its curvature increases substantially as \( |J_2| \to 0 \), indicating that the jump emerges as a singular limit for \( J_2 = 0 \). We note that Eqs.  ~\ref{eq:mh-dEgsdh} and ~\ref{eq:mh <>}  provide consistent results, further supporting our conclusion. Therefore, we expect other models with flat bands to exhibit similar features. Higher-order contributions in \( 1/S \) could alter these results, but such challenging calculations extend beyond the scope of this discussion.

It is tempting to relate this jump to the exact magnetization jump present for the \( S=1/2 \) case, which arises due to the localization of magnons ~\cite{schulenburg02, zhitomirsky04} and is linked to a \( m/S = 7/9 \) plateau, which has no semiclassical analogue.  Although both phenomena are rooted in the existence of a flat band in the kagome lattice, the semiclassical results are established only at order \( 1/S \), and we cannot predict how higher-order terms might affect them.  Nevertheless,  from general arguments, we expect that any finite $J_2$ will also destroy the jump in the quantum case \cite{capponi13,picot16}. This is because, upon acquiring a dispersion, the magnons delocalize, which means that the crystallization of a macroscopic number of them no longer occurs.  The \( 7/9 \) plateau does not immediately disappear due to its finite gap for \( h < h_c \).  Furthermore, Ref. \cite{zhitomirsky02} demonstrated that a finite temperature softens the jump. These two factors, when combined, make the experimental observation of the discontinuity challenging.  Overall, all results indicate that for small \( |J_2| \) and as \( T  \to 0 \), the magnetization curve should become very steep close to \( h_c \).

\section{\label{sec:conclusion}Conclusion}
We studied the spin-$S$ $J_1-J_2$ Heisenberg model in the kagome lattice and in the presence of an external magnetic field.  At zero field,  a finite $J_2$ lifts the massive degeneracy coming from the flat band,  with a positive (negative) $J_2$ selecting the $\vec{Q}=0$ ($\sqrt{3}\times\sqrt{3}$) state.  For a finite field,  however,  an accidental degeneracy persists,  which is lifted if we take into account the effects of thermal fluctuations,  an example of the order-by-disorder mechanism.  As fluctuations favor,  in general,  collinear states,  a plateau phase,  the up-up-down phase, is stabilized not only at $h_c/3$ --- where it is classically stable --- but over an extended field regime in analogy to what is observed in the triangular lattice ~\cite{chubukov91}.  Classical Monte Carlo results confirm this picture. 

To account for quantum fluctuations, we employed spin-wave theory to establish the existence of the plateau phase.  Already at the linear level,  the zero-point energy correction favors the plateau phase.  Moreover,  we showed that the suggestion of Ref.  ~\cite{coletta16} to construct the whole magnetization curve works also in our example: one can build this curve at the level of linear spin-wave by cutting the $1/S$ curves at $m/S=1/3$,  bypassing the need of investigating next-leading order,  in general.  We further justify these findings by using nonlinear spin waves in two complementary limits.  We showed that the pseudo-Goldstone modes associated with the accidental classical degeneracy acquire a gap in the collinear uud phase because of magnon-magnon interactions.  As long as the gap is finite,  the uud phase is stable.  The range of the uud phase can also be gauged by the stability of the neighboring Y and V phases by computing the $1/S$ correction to the classical angles.  To leading order in $1/S$, both nonlinear estimates agree.  However, our approach breaks down near $J_2=0$ because quantum fluctuations are too strong. We can estimate a region of stability for our spin-wave theory assuming that the $\delta S < S$,  where $\delta S$ is the $1/S$ correction to the order parameter.  For example, for $h=0$, our estimate for the absence of magnetic order is the region $-0.06 \lesssim J_2/J_1 \lesssim 0.05$ for spin $S=1/2$.

To stress the usefulness of the proposed phenomenological method, we compare its prediction of the plateau width with experimental and numerical data.  For example,  Ref ~\cite{kato24} found for a kapellasite-type compound the values of critical field for transitions to the plateau as $(h_1,h_2)=(0.8,1.6)$ in units of $J_1$, where our results for $J_2=0, S=1/2$ are $(h_1,h_2)=(0.865,1.675)$ in units of $J_1$. The critical field values broadly agree with other experiments \cite{jeon24,okuma19,suetsugu24,kermarrec21}, but with a caveat. Determining the strength of the exchange constants $J_1$ and $J_2$ is, for most compounds, not an easy task. Notable examples are compounds of the Herbertsmithite family, which have $J_1$ of the order of $100T$, meaning that the full magnetization curve up to saturation is not experimentally accessible. The practical route is that, when only $h_1$ is known, the approximation $h_1 \simeq h_c/3$ is used. Our results show that the melting of the plateau is always asymmetric, and the uud phase never penetrates too deeply into the Y phase, so this approach does not introduce large errors.  Unfortunately, even when both $h_1$ and $h_2$ can be measured, our estimate of the plateau width is largely insensitive to the strength of $J_2$.  It seems that the mechanism is basically dictated by $J_1$, and $J_2$ mainly selects the ordered state, either $\vec{Q}=0$ or $\sqrt{3}\times\sqrt{3}$.  Thus, using the plateau width alone to extract the values of the exchange constant produces insufficient information.  This suggests combining it with complementary information,  for instance, the Curie-Weiss temperature $\Theta_{CW}=S(S+1)z(J_1+J_2)/3k_B$, where $z=4$ is the number of neighbors. Note that for the case of $\vec{Q}=0$ spatial configuration,  $\Theta_{CW}$ can be directly related to the critical field $3k_B\theta_{CW}/(zS(S+1))=J_1+J_2=h_c/6S$, so we have prompt access to the important scale of the magnetization process.

One potential direction for future study is to incorporate additional interactions, such as the Dzyaloshinskii-Moriya interactions, which are relevant to the Herbertsmithite family ~\cite{norman16}. This interaction should, in principle, resist the selection of collinear phases and hence compete to destabilize the plateau, potentially leading to a rich phase diagram ~\cite{cepas01, nomura23}.  Additionally, it might be interesting to assess the effects of inhomogeneities on these plateau phases. On one hand,  defects can locally lift accidental degeneracies \cite{villain79}. On the other hand, we expect that they may destabilize noncollinear orders \cite{wollny12, santanu20, michel21, letouze25}, which could ultimately help stabilize the plateau.  Finally, understanding the connection between the current semiclassical description of the plateau --- valid for not too small \(J_2\) --- and the proposals to describe this phase as a magnon crystal is an intriguing open question.

%


\acknowledgments

We thank P.  C\^onsoli,  L.  Janssen,  and M. Vojta for discussions and collaborations on related topics.
We acknowledge support by FAPESP (Brazil),  Grants No. 2021/06629-4,  2022/15453-0,  and 2023/13425-1.  ECA was also supported by CNPq (Brazil), Grant No. 302823/2022-0.  


\appendix

\section{Details of the nonlinear spin-wave calculation} \label{sec:app_nlsw}

\begin{figure*}[t]
\centering{}\includegraphics[width=2\columnwidth]{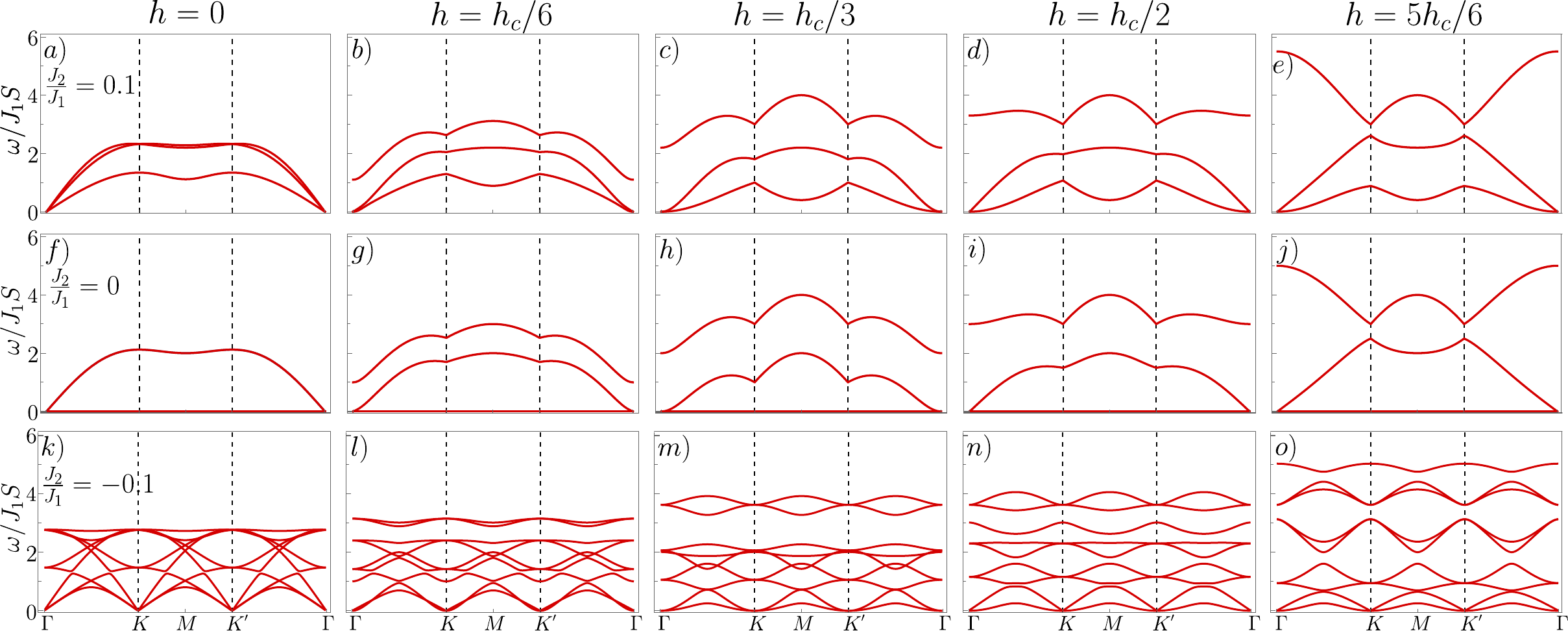}\caption{\label{fig:LSWT_Spect} Magnon dispersions $\omega_{\textbf{k}}$ for: (a)-(e) $J_2/J_1=0.1$; (f)-(j) $J_2/J_1=0$; and (k)-(o) $J_2/J_1=-0.1$; and and (a),  (f),  (k) $h=0$; (b),  (g),  (l) $h=h_c/6$; (c),  (h),  (m) $h=h_c/3$; (d),  (i),  (n) $h=h_c/2$; (e),  (j),  (o) $h=5h_c/6$; The critical field $h_c$ is $h_c=6J_1S$ for $J_2<0$ and $h_c=6(J_1+J_2)S$ for $J_2>0$. The path $\Gamma-K-M-K^{\prime}-\Gamma$ in the Brillouin zone is the same as that used in Fig. \ref{fig:pol_magnons}.  \label{fig:Spec_magnons} }
\end{figure*}

We start from a generic isotropic Heisenberg Hamiltonian in presence of a uniform magnetic field

\begin{equation}
    \mathcal{H}=\sum_{ij} J_{ij}\textbf{S}_i \cdot \textbf{S}_j-h\sum_i S_i^{z}.
\end{equation}
We work in the $S\to \infty$ limit and specialize in the case of coplanar spin configurations,  which can be parametrized by $\{\theta_i\}_{i=1,\dots N}$, where $N$ is the number of lattice sites.  We assume that the magnetic unit cell contains $N_s$ sites  and define $N_c$ as the total number of sublattices,  $N_c \equiv N/N_s$.  We perform rotations to local axes, such that the new spin operators in this basis are

\begin{equation}
\begin{split}
    S_{\mu i}^{z\prime}&=S_{\mu i}^{z}\cos\theta_{\mu}+S_{\mu i}^{x}\sin\theta_\mu\\ S_{\mu i}^{y\prime}&=S_{\mu i}^{y}\\ S_{ \mu i}^{z\prime}&=S^{x}_{\mu i}\sin\theta_{\mu}+S_{\mu i}^{z}\cos\theta_{\mu},
\end{split}
\end{equation}
where $\{\theta_{\mu}\}_{\mu=1,\dots N_s}$.  We apply the HP transformation to the bosonic operators $a_{\mu i}, a_{\mu i}^{\dagger}$. Taking a Fourier transform, we end up with
\begin{equation}
    \mathcal{H}=S^2E_{\rm{gs},0}-\frac{S}{2}\sum_{\textbf{k}}\rm{Tr}\mathbb{A}_{\textbf{k}}+\frac{S}{2}\sum_{\textbf{k}}\alpha_{\textbf{k}}^{\dagger}\mathbb{M}_{\textbf{k}}\alpha_{\textbf{k}},
\end{equation}
where $\alpha_{\textbf{k}}=(a_{\textbf{k},1}, a_{\textbf{k},2} \dots a_{\textbf{k},N_s},a_{-\textbf{k}1}^{\dagger},a_{-\textbf{k}2}^{\dagger},\dots.a_{-\textbf{k}N_s}^{\dagger})^{T}$ and

\begin{equation}
    \mathbb{M}_{\textbf{k}}=\begin{pmatrix}
        \mathbb{A}_{\textbf{k}} & \mathbb{B}_{\textbf{k}}\\
        \mathbb{B}_{\textbf{k}}^{\dagger} & \mathbb{A}^{T}_{-\textbf{k}}
    \end{pmatrix},
\end{equation}
with
\begin{equation}
\begin{split}
    \mathbb{A}^{\mu\nu}_{\textbf{k}}&=\delta^{\mu\nu}\frac{h^{\text{loc}}_{\mu}}{S}+\big[1+\cos\theta_{\mu\nu} \big]\mathbb{J}_{\mu\nu}(\textbf{k}), \\
    \mathbb{B}^{\mu\nu}_{\textbf{k}}&=\big[-1+\cos\theta_{\mu\nu}\big]\mathbb{J}_{\mu\nu}(\textbf{k)},
\end{split}
\end{equation}
where $\theta_{\mu\nu}=\theta_{\mu}-\theta_{\nu}$ and $\mathbb{J}_{\mu\nu}(\textbf{k})=\sum_{\nu j}\frac{J(\mu)i,(\nu)j}{2}e^{i\textbf{q}\cdot (\textbf{r}_{\nu j}-\textbf{r}_{\mu i})}$. Here, $h_{\mu}^{\text{loc}}$ is the local field at the $\mu$-th sublattice site,  defined as $h_{\mu}^{\text{loc}}=|h\hat{z}-\sum_{\nu j}J_{i(\mu)j(\nu)}\textbf{S}_j|$.  The transformation matrix $\mathbb{T}_{\textbf{k}}$ takes the HP bosons to new bosonic operators, $\beta_{\textbf{k}}=\mathbb{T}_{\textbf{k}}\alpha_{\textbf{k}}$, with $\beta_{\textbf{k}}=(b_{\textbf{k}1},b_{2\textbf{k}}, \dots,b_{\textbf{k}N_s},b_{-\textbf{k}1}^{\dagger},b_{-\textbf{k},2}^{\dagger},\dots,b^{\dagger}_{-\textbf{k}N_s})$ such that the Hamiltonian is diagonal in the $b_{\textbf{k}\mu}$'s.  The most general form of $\mathbb{T}_{\textbf{k}}$ is
\begin{equation}
    \mathbb{T}_{\textbf{k}}=\begin{pmatrix}
        \mathbb{X}_{\textbf{k}}^{*} & -\mathbb{Y}^{*}_{\textbf{k}}\\
        -\mathbb{Y}_{-\textbf{k}} & \mathbb{X}_{-\textbf{k}}
    \end{pmatrix}
\end{equation}
We have that the $b_{\textbf{k}\mu}$'s obey bosonic algebra if, and only if, $\mathbb{T}_{\textbf{k}}$ is pseudo-hermitian, that is,
$\sigma_3\mathbb{T}_{\textbf{k}}^{\dagger}\sigma_3 = \mathbb{T}^{-1}_{\textbf{k}}$,  where $\sigma_3\equiv\text{diag}(\mathbb{I}_{N_s},-\mathbb{I}_{N_s})$ ~\cite{blaizot1986}.  In that case, what we need to do is to diagonalize $\sigma_3\mathbb{M}_{\textbf{k}}$ and the $\mathbb{T}_{\textbf{k}}$'s are constructed with the eigenvectors obtained in the diagonalization. Some subtleties arise when dealing with $ 0$-eigenvalues; we refer the interested reader to Refs.~\cite {rau18,blaizot1986}. We end up with the Hamiltonian written as $\mathcal{H}=E_{gs}+S\sum_{\textbf{k}\mu}\omega_{\textbf{k}\mu}b^{\dagger}_{\textbf{k}\mu}b_{\textbf{k}\mu}$, with $E_{\text{gs}}$ given by Eq. \ref{eq:Enegs}. We plot the excitation energies $\omega_{\textbf{k}\mu}$ for some values of $h$ and $J_2$, see Fig. \ref{fig:Spec_magnons}. For $J_2 > 0$,  $\vec{Q}=0$ state,  we have three magnon bands,  whereas for $J_2 < 0$ we obtain nine magnon bands due to the enlarged magnetic unit cell in the $\sqrt{3} \times \sqrt{3}$ state.  The pseudo-Goldstone mode are present at the harmonic level,  they can be of type-I or type-II ~\cite{rau18},  highlighting the accidental degeneracy.  For $J_2=0$ we get the flat band at zero energy $\forall h$,  confirming the instability of the LSWT in this regime.

\subsection{Three-boson terms}

In terms of the classical parametrization, we have
\begin{equation}
\begin{split}
    S&\sqrt{S}\mathcal{H}_1=S\sqrt{\frac{S}{2}}\sum_{\mu i}(a_{\mu i}^{\dagger}+a_{\mu i})\big[ -\frac{h}{S}\sin\theta_{\mu}+\sum_{\nu j}J_{\mu i \nu j} \sin\theta_{\mu\nu}\big],\\
    &\sqrt{S}\mathcal{H}_3=\sqrt{\frac{S}{2}}\sum_{\mu i \nu j}J_{\mu i \nu j}(a_{\mu i}^{\dagger}+a_{\mu i})a_{\nu j}^{\dagger}a_{\nu j}\sin\theta_{\mu\nu}\\
    &-\frac{1}{4}\sqrt{\frac{S}{2}}\sum_{\mu i}(a_{\mu i}^{\dagger}a_{\mu i}^{\dagger}a_{\mu i}+h.c) \big[ -\frac{h}{S}\sin\theta_{\mu}+\sum_{\nu j}J_{\mu i \nu j}\sin \theta_{\mu \nu}\big].
\end{split}
\end{equation}
at harmonic order, it is required that $\mathcal{H}_{1}\equiv0$. This is automatically satisfied for a configuration that minimizes the classical energy, when
\begin{equation}
    \frac{\partial E_{\rm{gs},0}}{\partial\theta_{\mu}}=-\frac{h}{S}\sin\theta_{\mu}+\sum_{\nu j}J_{\mu i \nu j}\sin\theta_{\mu\nu}=0, \mu=1,\dots, N_s.
\end{equation}

Now we include the three- and four-boson terms by treating $\mathcal{H}_3$ and $\mathcal{H}_4$ perturbatively.  First, we use Wick's theorem ~\cite{molinari2023noteswickstheoremmanybody} to do a normal ordering with respect to the $b_{\textbf{k}\mu}$ bosons,
\begin{equation}
    \mathcal{H}_3=:\mathcal{H}_3:+\mathcal{H}_3^{(1)}
\end{equation}
where $\mathcal{H}^{(1)}_3$ are terms linear in bosons. Then, the condition for having no terms linear in bosons is to impose that $S\sqrt{S}\mathcal{H}_1(\tilde{\theta})+\sqrt{S}\mathcal{H}_3^{(1)}(\tilde{\theta})=0$, where $\tilde{\theta}=\theta+\delta\theta/S$. Note that for consistency is suficient to evaluate $\mathcal{H}_3^{(1)}$ at the classical $\theta$.
We define the averages

\begin{equation}
\begin{split}
    \Delta_{\mu i \nu j} &= \langle a_{\mu i}a_{\nu j} \rangle = \frac{1}{N_c}\sum_{\textbf{k}\lambda}e^{-i\textbf{k}\cdot(\textbf{r}_{\nu j}-\textbf{r}_{\mu j})}(\mathbb{X}_{\textbf{k}})_{\lambda \mu}(\mathbb{Y}_{\textbf{k}})^{*}_{\lambda \nu}\\
    m_{\mu i \nu j} &= \langle a^{\dagger}_{\mu i}a_{\nu j} \rangle =\frac{1}{N_c}\sum_{\textbf{k}\lambda}e^{-i\textbf{k}\cdot (\textbf{r}_{\nu j}-\textbf{r}_{\mu i})}(\mathbb{Y}_{\textbf{k}})_{\lambda \mu}(\mathbb{Y}_{\textbf{k}})^{*}_{\lambda \nu}\\
    n_{\mu}&= \langle a^{\dagger}_{\mu i}a_{\mu i}\rangle =  \frac{1}{N_c}\sum_{\textbf{k}\lambda}(\mathbb{Y}_{\textbf{k}})_{\lambda \mu}(\mathbb{Y}_{\textbf{k}})^{*}_{\lambda \mu}\\
    \delta_{\mu}&=\langle a_{\mu i}a_{\mu i}\rangle=\frac{1}{N_c}\sum_{\textbf{k}\lambda} (\mathbb{X}_\textbf{k})_{\lambda \mu}(\mathbb{Y}_{\textbf{k}})^{*}_{\lambda\mu}
\end{split}
\end{equation}
where the expected values are with respect to the vacuum of the $b_{\textbf{k}\mu}$ operators,  for $\mu=1,\dots,N_s$.  The renormalization condition consists of a linear system which can be solved for each $\delta\theta_{\mu}$.  In the present model,  the system may be underdetermined,  for both $\vec{Q}=0$ and $\sqrt{3}\times\sqrt{3}$ states.  To circumvent the resulting ambiguity,  we select the solutions that respect the configuration of the starting phase.  For example,  we choose $\delta\theta_2=-\delta\theta_3$ and $\delta\theta_1=0$ for the $Y$ phase in the $\vec{Q}=0$ state.  The resulting equations read

\begin{widetext}
        \begin{equation}
        \begin{split}
    -\frac{h}{S}&h^{(\text{loc})}_{\mu}\delta\theta_{\mu}-2\sum_{\nu}\mathbb{J}_{\mu\nu}(0)\cos\theta_{\mu\nu}\delta\theta_{\nu}
    =-\frac{2}{N_c}\sum_{\nu}\sin\theta_{\mu\nu}\sum_{\textbf{k}\lambda}\bigg\{\mathbb{J}_{\mu\nu}(\textbf{k})\big[(\mathbb{X}_{\textbf{k}})_{\lambda\nu}(\mathbb{Y}_{\textbf{k}})^{*}_{\lambda\mu}+(\mathbb{Y}_{\textbf{k}})_{\lambda\nu}(\mathbb{Y}_{\textbf{k}})^{*}_{\lambda\mu}\big]-\mathbb{J}_{\mu\nu}(0)(\mathbb{Y}_{\textbf{k}})_{\lambda\nu}(\mathbb{Y}_{\textbf{k}})^{*}_{\lambda\nu} \bigg\}
    \end{split}
    \end{equation}
    \label{eq:RenormLinearSystem}
\end{widetext}

\subsection{Four-boson terms}

Collecting all terms with four boson operators, we have
\begin{equation}
\begin{split}
      \mathcal{H}_4=\frac{1}{4}\sum_{ij}J_{ij}\bigg\{&a_i^{\dagger}a_ia_j^{\dagger}a_j \cos\theta_{ij}-\frac{1}{4}\big[ (-1+\cos\theta_{ij})a_ia_j^{\dagger}a_ja_j +\\
      &+(1+\cos\theta_{ij})a_ia_j^{\dagger}a^{\dagger}_{j}a_j + i \leftrightarrow j\big] + h.c. \bigg\}.
\end{split}
\end{equation}

Again, using Wick theorem we cast $\mathcal{H}_4$ in the normal ordered form, $\mathcal{H}^{(2)}_4= \; :\mathcal{H}_4:+:\mathcal{H}^{(2)}_4:+\mathcal{H}_4^{(0)}$. As discussed in Sec. \ref{sec:text_nlsw}, the relevant term in lowest order of $1/S$ is $\mathcal{H}^{(2)}_4$, which takes the explicit form (in momentum space) 

\begin{equation}
\begin{split}
        \mathcal{H}=\sum_{\textbf{k}}&\Big\{ \sum_{\mu}g_3^{\mu}a_{\textbf{k}\mu}^{\dagger}a_{\textbf{k}\mu}+ [g_4^{\mu}a_{\textbf{k}\mu}^{\dagger}a_{-\textbf{k}\mu}^{\dagger}+h.c.]\\
        &+\sum_{\mu\nu}[g_1^{\mu\nu}(\textbf{k})a_{\textbf{k}\mu}^{\dagger}a_{\textbf{k}\nu}+g_2^{\mu\nu}(\textbf{k})a_{\textbf{k}\mu}^{\dagger}a_{\textbf{k}\nu}^{\dagger}+h.c.]\Big\},
\end{split}
\end{equation}

where

\begin{widetext}
    \begin{equation}
\begin{split}
        g_1^{\mu\nu}&=\frac{1}{N_c}\Big\{ \sum_{\textbf{q}\lambda} (\mathbb{Y}_{\textbf{q}})_{\lambda \nu}(\mathbb{Y}_{\textbf{q}})^{*}_{\lambda\mu}\mathbb{J}_{\mu \nu}(\textbf{k}+\textbf{q})\Big\}\cos\theta_{\mu\nu} - \frac{1}{4}\Big\{(-1+\cos\theta_{\mu\nu})[\delta_{\mu}+\delta_{\nu}^{*}]+(1+\cos\theta_{\mu\nu})[n_{\nu}+n_{\mu}] \Big\}\mathbb{J}_{\mu\nu}(\textbf{k})\\
        g_2^{\mu\nu}&= \frac{1}{N_c}\Big\{\sum_{\textbf{q}\lambda} (\mathbb{X}_{\textbf{q}})_{\lambda \mu}(\mathbb{Y}_{\textbf{q}})^{*}_{\lambda \nu}\mathbb{J}_{\mu\nu}(\textbf{k}-\textbf{q}) \Big\}\cos\theta_{\mu\nu}-\frac{1}{4}\Big\{ (1+\cos\theta_{\mu\nu})[\delta_{\mu}+\delta_{\nu}]+(-1+\cos\theta_{\mu\nu})[n_{\mu}+n_{\nu}]\Big\}\mathbb{J}_{\mu\nu}(\textbf{k})\\
    g_3^{\mu}&=\sum_{\nu}2\mathbb{J}_{\mu\nu}(0)n_{\nu}\cos\theta_{\mu\nu}-\sum_{\nu}\frac{1+\cos\theta_{\mu\nu}}{2}\Big\{\frac{1}{N_c}\sum_{\textbf{q}\lambda}2\text{Re}\Big[(\mathbb{Y}_{\textbf{q}})_{\lambda \mu}(\mathbb{Y}_{\textbf{q}})^{*}_{\lambda \nu} \mathbb{J}_{\mu\nu}(-\textbf{q}) \Big]\Big\} -\\
    &-\sum_{\nu}\frac{-1+\cos\theta_{\mu\nu}}{2}\Big\{\frac{1}{N_c}\sum_{\textbf{q}\lambda} \Big[ (\mathbb{X}_{\textbf{q}})_{\lambda \mu}(\mathbb{Y}_{\textbf{q}})^{*}_{\lambda \nu}+(\mathbb{X}_{\textbf{q}})^{*}_{\lambda \nu}(\mathbb{Y}_{\textbf{q}})_{\lambda\mu}\Big]\mathbb{J}_{\mu\nu}(-\textbf{q})\Big\}\\
    g_4^{\mu}&=-\sum_{\nu}\frac{(1+\cos\theta_{\mu\nu})}{4}\Big\{ \frac{1}{N_c}\sum_{\textbf{q}\lambda} (\mathbb{X}_{\textbf{q}})_{\lambda \mu} (\mathbb{Y}_{\textbf{q}})^{*}_{\lambda\nu}\mathbb{J}_{\mu\nu}(-\textbf{q})\Big\}-\sum_{\nu}\frac{(-1+\cos\theta_{\mu\nu})}{4}\Big\{\frac{1}{N_c}\sum_{\textbf{q}\lambda}(\mathbb{Y}_{\textbf{q}})_{\lambda \nu}(\mathbb{Y}_{\textbf{q}})^{*}_{\lambda \mu}\mathbb{J}_{\mu\nu}(\textbf{q}) \Big\}
\end{split}
\end{equation}
\end{widetext}

This can be written in the convenient form,
\begin{equation}
    \mathcal{H}_4^{(2)}=\frac{1}{2}\sum_{\textbf{k}}\alpha_{\textbf{k}}^{\dagger}\tilde{\mathbb{M}}_{\textbf{k}}\alpha_{\textbf{k}} - \frac{1}{2}\sum_{\textbf{k}}\text{Tr}\;\tilde{\mathbb{A}}_{\textbf{k}}
    \label{eq:H4matrixform}
\end{equation}

where $\alpha_{\textbf{k}}=(a_{\textbf{k},1}\dots a_{\textbf{k}N_s},a_{-\textbf{k}1}^{\dagger},\dots,a_{-\textbf{k}N_s}^{\dagger})^{T}$ and

\begin{equation}
    \tilde{\mathbb{M}}_{\textbf{k}}=\begin{pmatrix}
        \tilde{\mathbb{A}}_{\textbf{k}} & \tilde{\mathbb{B}}_{\textbf{k}}\\
        \tilde{\mathbb{B}}_{\textbf{k}}^{\dagger} & \tilde{\mathbb{A}}_{-\textbf{k}}^{T}
    \end{pmatrix},   
\end{equation}
with $\tilde{\mathbb{A}}_{\textbf{k}}^{\mu\nu}=g_3^{\mu}\delta_{\mu\nu} + 2g_1^{\mu\nu}(\textbf{k)} \;, \tilde{\mathbb{B}}_{\textbf{k}}^{\mu\nu}=2g_4^{\mu}\delta_{\mu\nu}+2g_2^{\mu\nu}(\textbf{k})$. The last term in Eq. \ref{eq:H4matrixform} amounts to a momentum-independent shift of the ground-state energy. We treat the first term on the right-hand side using perturbation theory. If $\ket{\textbf{k}\mu}\equiv b_{\textbf{k}\mu}^{\dagger}\ket{0}$ then the correction of the energy excitations is $\delta \omega_{\textbf{k}\mu}=\bra{\textbf{k}\mu}\frac{1}{2}:\alpha_{\textbf{k}}^{\dagger}\tilde{\mathbb{M}}_{\textbf{k}}\alpha_{\textbf{k}}:\ket{\textbf{k}\mu}=\frac{1}{2}\bra{\textbf{k}\mu}:\beta^{\dagger}_{\textbf{k}}(\mathbb{T}_{\textbf{k}}^{-1})^{\dagger}\tilde{\mathbb{M}}_{\textbf{k}}\mathbb{T}_{\textbf{k}}^{-1}\beta_{\textbf{k}}:\ket{\textbf{k}\mu}=\frac{1}{2}(\Sigma^{\mu\mu}_{\textbf{k}}+\Sigma_{-\textbf{k}}^{\mu+N_s,\mu+N_s})$.


%

\end{document}